\documentclass[sigconf]{acmart}

\usepackage{booktabs} 

\setcopyright{rightsretained}

\acmDOI{}

\acmISBN{}

\acmConference[PCSC2020]{Philippine Computing Science Congress}{March 2020}{Baguio City, Philippines}
\acmYear{2020}
\copyrightyear{2020}

\acmArticle{}
\acmPrice{}

\editor{}
\editor{}
\editor{}

\begin{document}
\title{Investigation of Dataset Features for Just-in-Time Defect Prediction}

\author{Giuseppe Ng}
\orcid{0000-0002-1375-0236}
\affiliation{%
  \institution{De La Salle University}
  \streetaddress{2401 Taft Ave.}
  \city{Malate}
  \state{Manila}
  \postcode{1004}
}
\email{giuseppe_ng@dlsu.edu.ph}

\author{Charibeth Cheng}
\affiliation{%
  \institution{De La Salle University}
  \streetaddress{2401 Taft Ave.}
  \city{Malate}
  \state{Manila}
  \postcode{1004}
}
\email{charibeth.cheng@dlsu.edu.ph}

\begin{abstract}
Just-in-time (JIT) defect prediction refers to the technique of predicting whether a code change is defective. Many contributions have been made in this area through the excellent dataset by Kamei. In this paper, we revisit the dataset and highlight preprocessing difficulties with the dataset and the limitations of the dataset on unsupervised learning. Secondly, we propose certain features in the Kamei dataset that can be used for training models. Lastly, we discuss the limitations of the dataset's features.
\end{abstract}

%
%



\maketitle

\section{Introduction}
Over time, the use and reliance to systems have increased and has lead to an exponential increase in complexity and size \cite{Ghose2018,Li2018}. This increase of complexity through new functionalities also introduce more defects \cite{Nam2018}. As such, the impact of software defects have also increased \cite{Ge2018,Huang2018}.

Automated tools can assist software engineers in identifying areas that contain flaws so that developers can focus optimization efforts and projects testing costs can be reduced. This can be critical especially for smaller software companies with a limited testing budget \cite{Ghose2018,Li2018,Son2019}. Software Defect Prediction (SDP) hence pertains to the detection of detect potential defects \cite{Qiao2019}. The goal of defect prediction is to identify the faults in the code and code modifications, estimate the number of defects and identify areas of optimization \cite{Hosseini2019}.

Traditional SDP approaches treat this challenge as a binary classification: defective or non-defective \cite{Qiao2019}. Simple prediction modeling techniques have been adopted in the field such as  KNN, Naive Bayes, Support Vector Machine (SVM) and decision trees such as Random Forests \cite{Ge2018,Li2018}. Supervised models are expected to learn from a corpus of data that encompasses code archive in classes or files form and source trees such as GIT, SVN, or CVS, and defect information in a bug ticketing system such as Bugzilla or Jira \cite{Hosseini2019,Li2018}. Linking source commit information and bug tickets may also be necessary. Granularity of defect prediction can be by package, class, file, or function basis \cite{Hosseini2019,Qiao2019}.

Majority of the works on SDP focus on detecting defective modules or files, however, these are coarse grained approaches, leaving the work of discovering actual faulty lines to the developers \cite{Kamei2013}. Mockus and Weiss \cite{Mockus2000} were first in proposing to identifying faulty code changes or commits instead of files, classes or modules \cite{Huang2017}. This technique is referred to as \textit{Just-in-Time Defect Prediction}. Recently, this area of research has become prevalent due to its time-sensitive nature and its fine-grained potential in defect prediction \cite{82ac60c1c56b4be4a6220de0dfad6039}. This approach to defect prediction can be useful when checking in new code, reducing the number of lines inspected by a developer \cite{Huang2017,Kamei2013}. Considering the lines of code to be inspected is referred to as Effort-aware Just-in-Time defect prediction \cite{Qiao2019}.

Unsupervised learning has also been proposed, most notably by \cite{Yang2016}. Unsupervised learning attempts to address the difficulty in labeling training data \cite{Huang2017,Yang2016}. In comparing the supervised and unsupervised approaches, Huang et al. \cite{Huang2017} concluded that the simple unsupervised models proposed in \cite{Yang2016} improved recall while sacrificing precision due to false positives. Yang's work \cite{Yang2016} is further refuted by \cite{Fu2017}.

Several works in Effort-aware Just-in-Time defect prediction utilize the main dataset provided by Kamei and propose utilizing the same features for their experiments \cite{Fu2017,Kamei2013,Qiao2019,Yang2016}. As pointed in \cite{82ac60c1c56b4be4a6220de0dfad6039}, more research work is needed to enhance the potential of Just-in-Time defect prediction.

In our feature analysis of the dataset, we report on some notable challenges with preprocessing that researchers may encounter in the dataset from \cite{Kamei2013} and share our findings on the most important features through Principal Component Analysis (PCA). We validate the features through the use of a simple neural network.

\section{Background}
In this section, we discuss the details of Kamei's dataset and the feature extraction, the preprocessing techniques, and the basic process of building a classifier \cite{Huang2018,Kamei2013,Qiao2019}. 

\subsection{Kamei's Dataset}
The dataset is created using six open source projects: (1) Bugzilla, (2) Columba, (3) Eclipse JDT, (4) Eclipse Platform, (5) Mozilla, and (6) Postgres SQL. The Bugzilla and Mozilla  datasets were generated using data provided by Mining Software Repository 2007 Challenge\footnote{\href{http://2007.msrconf.org/challenge/}{http://2007.msrconf.org/challenge/}}. The Eclipse JDT and Platform data were provided by the MSR 2008 Challenge\footnote{\href{http://2008.msrconf.org/challenge/}{http://2008.msrconf.org/challenge/}}. Columba and Postgres were from their respective CVS repositories \cite{Kamei2013}.

Labeling defective changes is done through the SZZ algorithm \cite{Sliwerski2005}. An open source implementation was created by \cite{Borg2019}. With regards to the Columba and Postgres dataset, an approximated SZZ approach was taken due to the defect ids not referenced in the commit messages \cite{Kamei2013}.

Table \ref{tab:datasettable} illustrates the features of the datasets \cite{Kamei2013}.

\begin{table}[htbp]
\caption{Features of the Kamei dataset \cite{Fu2017,Kamei2013}}
\begin{center}
\begin{tabular}{cc}
\toprule
\textbf{Metric} & \textbf{\textit{Description}} \\
\midrule
NS & Number of modified subsystems  \\
ND & Number of modified directories  \\
NF & Number of modified files  \\
Entropy & Distribution of the modified \\
& code across each file  \\
\hline
LA & Lines added  \\
LD & Lines deleted  \\
LT & Lines of code in a file before the current change \\
\hline
Fix & Whether the change is a bug fix \\
\hline
NDEV & Number of developers that changed the modified file \\
AGE & The average time interval between the last \\
 & and the current change \\
NUC & The number of unique changes to the modified file \\
\hline
EXP & Developer experience in terms of number of changes \\
REXP & Recent developer experience \\
SEXP & Developer experience on the subsystem \\
\bottomrule
\end{tabular}
\label{tab:datasettable}
\end{center}
\end{table}

As discussed by Kamei \cite{Kamei2013}, the diffusion aspect is represented by NS, ND, NF and Entropy. Highly distributed changes are more complex and are more likely to be defective. Number of modified subsystems, modified directories, and modified files using the root directory name, directory names, and file names respectively \cite{Kamei2013}.

\begin{figure}[htbp]
\centerline{\includegraphics[width=0.5\textwidth]{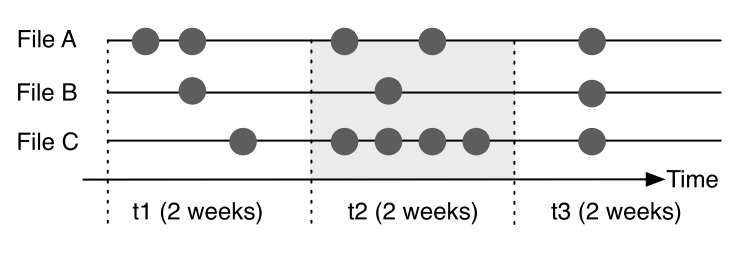}}
\caption{Example of Entropy \cite{DAmbros2010}.}
\label{fig:entropy}
\end{figure}

Entropy is calculated using a similar measure to Hassan \cite{Hassan2009,Kamei2013}. It is the measurement over time how distributed the changes are in the code base. Changes in one file is simpler than one that impacts many different files. Time period used for the dataset is 2 weeks \cite{Kamei2013}. Entropy is defined as \cite{Kamei2013}:

\begin{equation}
H(P)=- \sum_{k=1}^{n}(p_k*log_2 p_k) \label{eq:entropy}
\end{equation}

where probabilities $p_k \geq 0, \forall k \in 1,2, ..., n$, $n$ is the number of files in the change, $P$ is a set of $p_k$, where $p_k$ is the proportion that $file_k$ is modified in a change and $(\sum_{k=1}^{n} p_k) = 1$ \cite{Kamei2013}. 
Figure \ref{fig:entropy} illustrates entropy in a sample scenario \cite{DAmbros2010}.

For time interval $t1$, there are four changes and the probabilities are $P_A = \frac{2}{4}, P_B = \frac{1}{4}, P_C = \frac{1}{4}$. The entropy in $t1$ is calculated as $H = -(0.5 * log_2 0.5 + 0.25 * log_2 0.25 + 0.25 * log_2 0.25) = 1$. For time interval $t2$, entropy is 1.378 \cite{DAmbros2010}.

Lines added (LA), lines deleted (LD), and lines of code before the current change (LT) were calculated directly from the source code. LA and LD were normalized by dividing by LT while LT is normalized by dividing by NF since these two metrics have high correlation \cite{Kamei2013}. NUC is calculated by counting the number of commits that caused changes to specific files. This metric was also normalized by dividing by NF due to their correlation with the NF feature \cite{Kamei2013}. This technique follows findings made by Nagappan and Ball \cite{Nagappan2005}.

Age is calculated as the average of time interval between the last and current change of files \cite{Fu2017,Kamei2013}. Should Files A, B and C were modified 3 days ago, 5 days ago, and 4 days ago respectively, Age is calculated as 4 (i.e., $\frac{3+5+4}{3}$). More detailed explanation of the features of the dataset can be found in \cite{Kamei2013}.

The table \ref{tab:datasetsummary} shows the brief summary of the datasets that were generated in \cite{Kamei2013}.

\begin{table}[htbp]
\caption{Summary of the datasets \cite{Kamei2013,Yang2016}}
\begin{center}
\begin{tabular}{cccc}
\toprule
 &  &  & \textbf{\%Defect} \\
 &  &  & \textbf{induced} \\
\textbf{Project} & \textbf{Period} & \textbf{\#change} & \textbf{change} \\
\midrule
BUG & 08/1998-12/2006 & 4620 & 36\% \\
COL & 11/2002-07/2006 & 4455 & 14\% \\
JDT & 05/2001-12/2007 & 35386 & 14\% \\
PLA & 20/2001-12/2007 & 64250 & 5\% \\
MOZ & 01/2000-12/2006 & 98275 & 25\% \\
POS & 07/1996-05/2010 & 20431 & 20\% \\
\bottomrule
\end{tabular}
\label{tab:datasetsummary}
\end{center}
\end{table}

\subsection{Features Used in Existing Studies}
NS, NM, NF, NDEV, PD, EXP, REXP and SEXP are raw data. Entropy, LA, LD, LT and NUC are normalized. Fix is a boolean value \cite{Qiao2019}. In the downloaded dataset from Kamei \cite{Kamei2013}, NM refers to ND, PD refers to AGE, and NPT refers to NUC.

In \cite{Huang2017}, ND and REXP are removed because NF and ND are correlated. Correlated features will decrease the accuracy of the classifier. LA and LD were also found to be highly correlated \cite{Kamei2013}. In addition, LA and LD are not included as learning features because these are used to sort the inspection effort \cite{Huang2017}. 

The feature recommendations when using Just-in-Time defect prediction from these studies are: NS, NF, Entropy, LT, FIX, NDEV, PD, NPT, EXP, SEXP. Given the Just-in-Time defect prediction $Y(x)$, effort-aware assessment is calculated as $\frac{Y(x)}{Effort(x)}$ where $Effort(x) = LA_x + LD_x$ \cite{Kamei2013,Yang2016}.These insights were echoed in \cite{Qiao2019}. 

\subsection{Dataset Preprocessing}
Two specific issues with the dataset were highlighted in \cite{Kamei2013}. First, the dataset is highly skewed \cite{Huang2017}. This is addressed through the use of logarithmic transformation \cite{Kamei2013,Qiao2019}.

Data imbalance is another issue where in the defect-inducing changes only make up a small percentage of all changes in a project. This can be addressed by performing random undersampling. Undersampling, however, is not done on the validation set \cite{Kamei2013}.

\subsection{Building the Classifier}
The process of building the classifier begins with resampling the training set through undersampling \cite{Kamei2013}. While oversampling can be used, \cite{Huang2018} found that undersampling was more effective \cite{Bennin2018,Hosseini2019}. After, ND, REXP, LA, and LD are removed from the features of the training set. Then, a standard logarithm transformation is applied on each of the metrics. The classifier can then be trained \cite{Huang2018}.

\subsection{Principal Component Analysis}
Principal Component Analysis (PCA) is a technique used to reduce the complex features into simple principal components that are seen as the summary of the features. It is an unsupervised learning technique used to identify patterns and clusters from a particular dataset \cite{Lever2017}. It can be used to discover hidden features that are not readily seen in analytical studies \cite{Bo2008}. To the best of our knowledge, we have not seen PCA applied into the Kamei Dataset. We use this technique to uncover hidden attributes with this dataset.

\section{Research Methodology}
The dataset are examined to identify any visible issues. PCA is done on the raw dataset to gain a visual analysis of the dataset and the identify potential recommended features. Initially, no preprocessing is done to ensure that the test is easily reproduceable. PCA is performed on all the datasets and the distinct features found will be selected for experimentation.

Following the process of building the classifier, we perform preprocessing on the dataset and report on the decisions we made. For validation, we conduct experiments using a simple neural network on the feature selection and other feature combinations. The evaluation of the features selected is based on the Recall. These metrics are typically used in evaluating classifiers in SDP \cite{Alsaeedi2019,Wang2010}. The equation is shown below \cite{Xu-YingLiu2009}:
\begin{equation}
	Recall = \frac{TP}{TP+FN}
\end{equation}

Where TP is the True Positive, and FN is the False Negative. Final recall reported is the average of all recall tests of the model.

\section{Research Findings}
In this section, we discuss the findings of our dataset study, PCA and experiments.

\subsection{Non-normalized Instances Found}
In analyzing the features of the dataset in \cite{Kamei2013}, we found that usually, when LT is 0, LA and LD are also 0. However, there are also instances where LA and LD are not 0. This most likely means that the submitted changes consisted of completely new files. In such a scenario, LA and LD are not normalized and instead retain their raw data. LA and LD are usually only factored into Effort-Aware sorting and we wonder what impact these characteristics could have. Though these instances are few, in new software projects, there are more instances of new files being added into a source repository.

\subsection{Pre-processing Challenges}
Logarithm transformation is done to address skewness of data \cite{Kamei2013}. However, some of the data in the features lead to $log(0)$ which can cause infinity values in the dataset. To the best of our knowledge, we have not found a paper that raises this concern. The work by \cite{Bellego2019} discusses in general this issue though it is not specific to the field of defect prediction.

In \cite{Qiao2019}, normalization of data is suggested using the equation below:
\begin{equation}
	Norm(x) = \frac{x - min(x)}{max(x)-min(x)}
\end{equation}
However, in light that some instances of LA and LD are raw values, we wonder what the impact of this is.

\subsection{Overlapping of Non-defective and Defective Dataset}
Visualizing the dataset as shown in earlier figures indicate that the challenge in training for Just-in-Time defect prediction is that the labeled data for defective and non-defective changes overlap tightly. As unsupervised learning involves clustering according to \cite{Lever2017}, the overlapping nature of the PCA shows the limitation of Kamei's dataset with regards to unsupervised learning. This is consistent with the findings of \cite{Huang2017} and \cite{Yan2017} where unsupervised learning methods do not perform well compared to supervised learning methods.

\subsection{Analyzing Kamei's Recommended Features}
\begin{figure}[htbp]
\centerline{\includegraphics[width=0.5\textwidth]{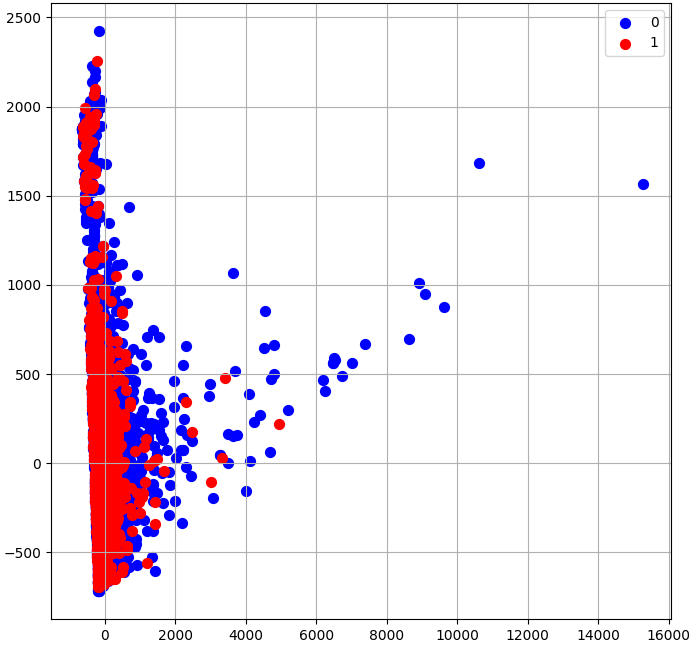}}
\caption{PCA of Bugzilla with all the Features of the Dataset.}
\label{fig:bugz-all-feat}
\end{figure}

Using the principal component analysis (PCA) technique, we compare the PCA of the Bugzilla dataset with all the features and the recommended features. Figure \ref{fig:bugz-all-feat} illustrates this. Blue refers to non-defective while red indicates defective change.

\begin{figure}[htbp]
\centerline{\includegraphics[width=0.5\textwidth]{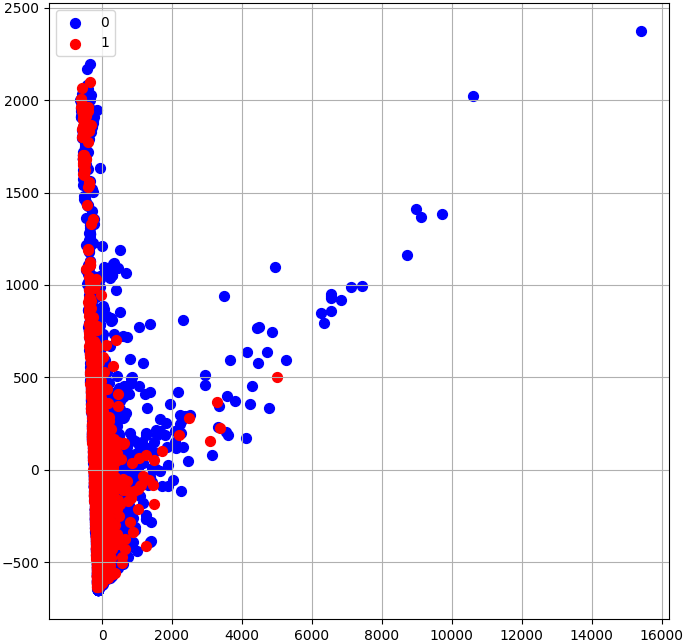}}
\caption{PCA of Bugzilla with Kamei's Recommended Features.}
\label{fig:bugz-rec-feat}
\end{figure}

Figure \ref{fig:bugz-rec-feat} illustrates the PCA with the recommended features from \cite{Kamei2013}, namely, NS, NF, Entropy, LT, FIX, NDEV, PD, NPT, EXP, SEXP. Observations show that there is minimal change on the dataset.

This suggests that using the recommended features do not alter the behaviour of the dataset for training purposes. Surprisingly, however, this PCA can be achieved with close proximity by utilizing only two features in the Bugzilla dataset: LT, and PD.

\begin{figure}[htbp]
\centerline{\includegraphics[width=0.5\textwidth]{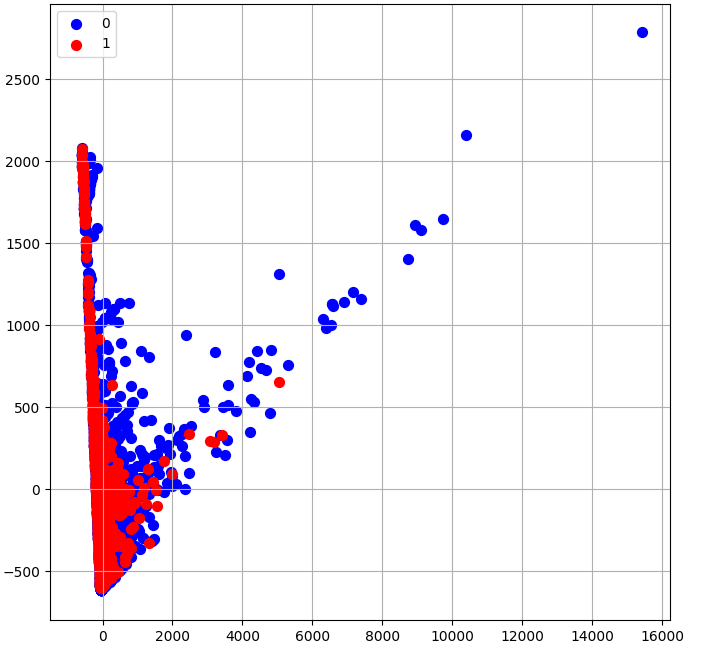}}
\caption{PCA of Bugzilla using LT and PD.}
\label{fig:bugz-pdlt-feat}
\end{figure}

Interestingly, these features were highlighted in the resulting unsupervised model by \cite{Yang2016}.

\begin{figure}[htbp]
\centerline{\includegraphics[width=0.5\textwidth]{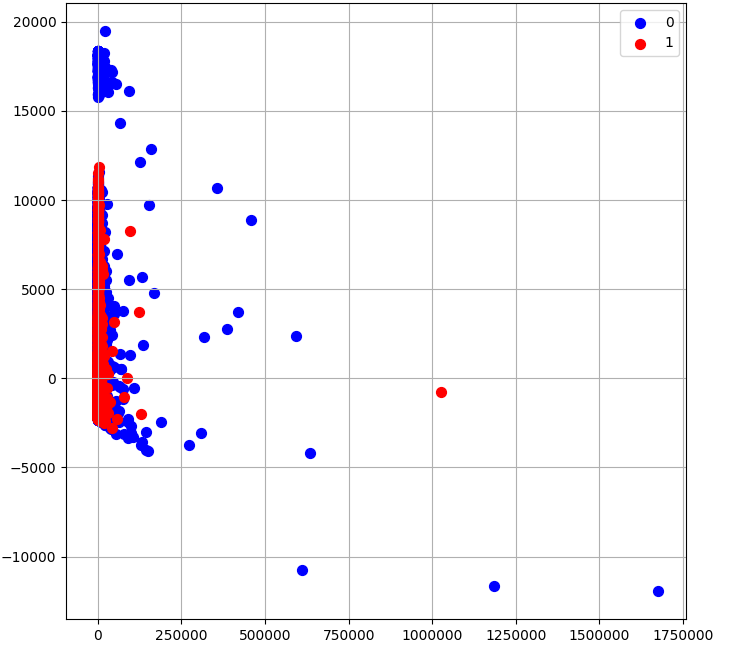}}
\caption{PCA of Mozilla using Kamei's Recommended Features.}
\label{fig:moz-rec-feat}
\end{figure}

\begin{figure}[htbp]
\centerline{\includegraphics[width=0.5\textwidth]{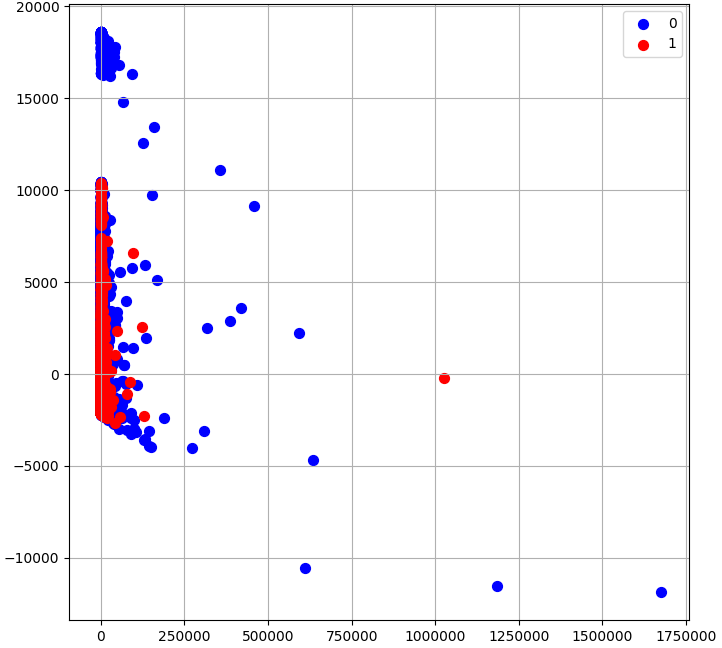}}
\caption{PCA of Mozilla using EXP and PD.}
\label{fig:moz-exp-pd-feat}
\end{figure}

In the Mozilla dataset, Figures \ref{fig:moz-rec-feat} and \ref{fig:moz-exp-pd-feat} illustrate the PCA visualizations of recommended features and EXP, PD respectively.

\begin{figure}[htbp]
\centerline{\includegraphics[width=0.5\textwidth]{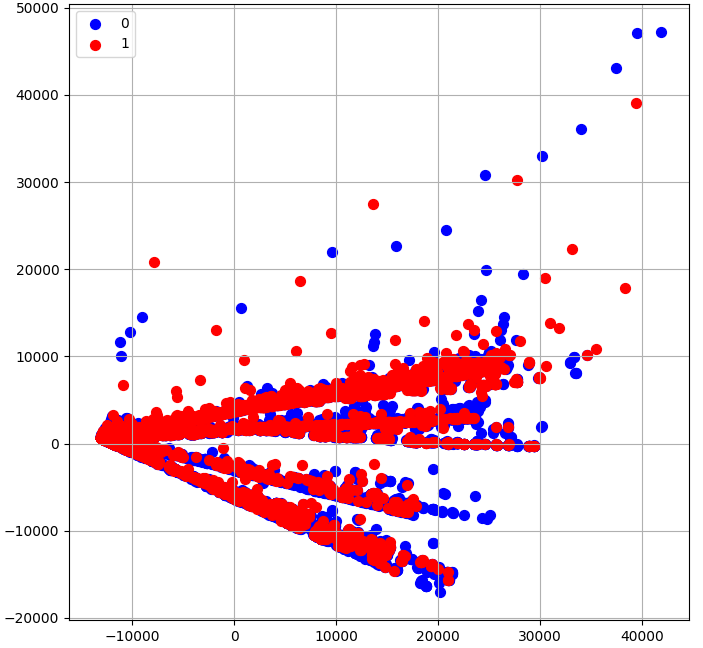}}
\caption{PCA of Postgres using EXP, SEXP and PD.}
\label{fig:pos-exp-sexp-pd-feat}
\end{figure}

The PCAs for Columba, Eclipse JDT and Platform datasets also point to EXP and PD. Interestingly, the Postgres dataset showed that PD, EXP, and SEXP. Figure \ref{fig:pos-exp-sexp-pd-feat} shows the PCA for Postgres.

With these results from PCA, we focus on using the following features: (1) LT, (2) PD, (3) EXP, (4) SEXP. LT as an important feature corroborates with the findings of \cite{Moser2008} and \cite{Nagappan2005}. The selection of PD is consistent with the work of \cite{Kamei2013} and \cite{Matsumoto2010}. EXP and SEXP being important features is consistent with the findings in \cite{Matsumoto2010} and \cite{Mockus2000} where more experienced developers are less likely to produce defective code. Experience-related features can be used in defect prediction according to \cite{Pascarella2019}.

Given that the Bugzilla, Mozilla and Postgres datasets yielded the unique set of features. We use these datasets in training our model with the features identified in the PCA. Our experiments focus on these identified features to determine which are most important.

In terms of validation, the neural network design used is inspired by \cite{Qiao2019} with some slight deviations. Our implementation is on PyTorch and uses 2 hidden layers (20 tanh activation nodes, 10 ReLU nodes) and a Linear output layer. Epoch used was 150, with an Adam Optimizer and a learning rate of 0.001 with a drop out layer ($p=0.2$). 10-fold cross validation on our training. The Bugzilla dataset was selected with a training set, testing set split of 90:10. Effort aware computations were not considered as our focus was on the regression model itself and that the features useful for effort aware prediction were not part of the features considered during PCA.

We used the features recommended by previous works earlier and also the LT, PD combination to see whether the results were comparable. The following experiment was done using the Bugzilla dataset whose training data has been randomly undersampled, natural logarithm applied on NS, NF, NDEV, NPT, EXP, REXP, and SEXP. Normalization performed on all features. These are the preprocessing steps advised by \cite{Qiao2019} though we limited logarithm operations to these columns to avoid the infinity values from entering the dataset. For evaluation, we used recall as suggested in \cite{Qiao2019}. Table \ref{tab:experimentqiao} shows the results on the Bugzilla dataset.

\begin{table}[htbp]
\caption{Experiment on Bugzilla Dataset}
\begin{center}
\begin{tabular}{cc}
\toprule
\textbf{Features} & \textbf{Recall} \\
\midrule
LT,PD & 56.80\% \\
LT,PD,SEXP & 55.17\% \\
PD,SEXP & 54.20\% \\
LT,PD,Entropy,NS,NF,Fix & 54.04\% \\
LT,PD,Entropy,NS,NF,Fix,NDEV & 50.32\% \\
LT,PD,Entropy,NS,NF,Fix,NDEV,NPT & 48.59\% \\
LT,PD,Entropy & 46.35\% \\
LT,PD,Entropy,NS & 44.98\% \\
LT,PD,Entropy,NS,NF & 41.69\% \\
LT,PD,Entropy,NS,NF,Fix,NDEV,NPT,EXP & 40.46\% \\
LT,PD,Entropy,SEXP & 38.30\% \\
PD,EXP,SEXP & 34.95\% \\
NS,NF,Entropy,LT,FIX,NDEV,PD,NPT,EXP,SEXP & 33.68\% \\
LT,PD,EXP & 17.28\% \\
PD,EXP & 6.89\% \\
\bottomrule
\end{tabular}
\label{tab:experimentqiao}
\end{center}
\end{table}

The LT,PD model and LT,PD,SEXP model performed notably better than LT,PD,EXP and LT,PD,SEXP,Entropy. This suggests that as far as the Bugzilla dataset is concerned, LT,PD is the minimal most influential features for consideration. 

Table \ref{tab:experimentqiao-mozilla} shows the results when using a Mozilla dataset. Tests using the Postgres dataset are shown on Table \ref{tab:experimentqiao-postgres}.

\begin{table}[htbp]
\caption{Experiment on Mozilla Dataset}
\begin{center}
\begin{tabular}{ccc}
\toprule
\textbf{Features} & \textbf{Recall} \\
\midrule
LT,PD,Entropy,NS,NF & 71.36\% \\
LT,PD,SEXP & 69.50\% \\
LT,PD & 66.69\% \\
PD,EXP,SEXP & 63.49\% \\
LT,PD,Entropy,NS,NF,Fix,NDEV,NPT,EXP & 63.25\% \\
LT,PD,Entropy & 60.76\% \\
LT,PD,Entropy,NS & 60.36\% \\
LT,PD,Entropy,NS,NF,Fix,NDEV & 57.50\% \\
LT,PD,Entropy,SEXP & 57.48\% \\
PD,SEXP & 56.77\% \\
LT,PD,Entropy,NS,NF,Fix & 56.83\% \\
LT,PD,Entropy,NS,NF,Fix,NDEV,NPT & 55.04\% \\
NS,NF,Entropy,LT,FIX,NDEV,PD,NPT,EXP,SEXP & 51.14\% \\
PD,EXP & 14.72\% \\
LT,PD,EXP & 12.24\% \\
\bottomrule
\end{tabular}
\label{tab:experimentqiao-mozilla}
\end{center}
\end{table}

\begin{table}[htbp]
\caption{Experiment on Postgres Dataset}
\begin{center}
\begin{tabular}{ccc}
\toprule
\textbf{Features} & \textbf{Recall} \\
\midrule
LT,PD & 65.85\% \\
LT,PD,Entropy,NS & 62.20\% \\
LT,PD,Entropy & 61.58\% \\
LT,PD,Entropy,SEXP & 61.51\% \\
LT,PD,Entropy,NS,NF & 58.85\% \\
LT,PD,SEXP & 58.18\% \\
PD,SEXP & 51.78\% \\
LT,PD,EXP & 41.82\% \\
LT,PD,Entropy,NS,NF,Fix,NDEV,NPT & 38.63\% \\
PD,EXP & 32.34\% \\
LT,PD,Entropy,NS,NF,Fix,NDEV & 32.01\% \\
PD,EXP,SEXP & 30.82\% \\
LT,PD,Entropy,NS,NF,Fix,NDEV,NPT,EXP & 28.20\% \\
LT,PD,Entropy,NS,NF,Fix & 28.15\% \\
NS,NF,Entropy,LT,FIX,NDEV,PD,NPT,EXP,SEXP & 23.98\% \\
\bottomrule
\end{tabular}
\label{tab:experimentqiao-postgres}
\end{center}
\end{table}

With the test done, despite the fact that EXP,PD was the deduced features influencing the PCA for Mozilla, LT,PD performed far better. Interestingly, these features performed better than the features recommended that was reported in previous works.

LT,SEXP,PD,Entropy only performed well in the Mozilla dataset while doing significantly worse in the Bugzilla dataset. \\LT,PD,Entropy,NS,NF was surprisingly good in the Mozilla dataset though it did not perform nearly as well on the Bugzilla dataset. LT,PD, PD,SEXP, and LT,PD,SEXP performed fairly closely in the datasets that we ran our experiments on. PD,EXP,SEXP performed well in the Mozilla dataset.

We decided to select notable features based on these results and run a different test on a combined dataset of Bugzilla and Mozilla to see a comparison of their performance on a modified dataset. The combined dataset utilized the preprocessed Bugzilla and Mozilla datasets with undersampling for equal distribution of classes between the two target datasets. Table \ref{tab:experimentqiao-bigtwo} shows the results.

\begin{table}[htbp]
\caption{Selected Feature Combinations on Combined Bugzilla and Mozilla Dataset}
\begin{center}
\begin{tabular}{ccc}
\toprule
\textbf{Features} & \textbf{Recall} \\
\midrule
LT,PD & 69.03\% \\
LT,PD,SEXP & 66.55\% \\
PD,SEXP & 59.09\% \\
LT,PD,Entropy,NS,NF & 58.34\% \\
LT,PD,Entropy,SEXP & 58.26\% \\
LT,PD,Entropy,NS,NF,Fix,NDEV,NPT,EXP & 55.88\% \\
NS,NF,Entropy,LT,FIX,NDEV,PD,NPT,EXP,SEXP & 47.11\% \\
PD,EXP,SEXP & 35.63\% \\
\bottomrule
\end{tabular}
\label{tab:experimentqiao-bigtwo}
\end{center}
\end{table}

LT,PD performs well with LT,PD,SEXP close to this feature. LT,PD seemed to be a reliable feature set to use in the dataset. With the results shown of our experiments, SEXP is a possible complement to LT and PD. The other features did not stand out in all of the experiments.

\section{Analysis}
Based on our own findings, LT and PD are important features. This is different from findings from \cite{Kamei2013} where the features recommended were NS, NF, Entropy, LT, FIX, NDEV, PD, NPT, EXP, SEXP. Our selection of LT as a notable feature is in disagreement with the results by \cite{Pascarella2019}. However, LT and PD were the features recommended by \cite{Yang2016} in their work on unsupervised models. The result from \cite{Matsumoto2010} and \cite{Mockus2000} indicate experience as an important dimension in determining defective changes. Experience-based metrics was also mentioned as good features in \cite{Pascarella2019}. In our experiment, feature combinations with EXP did not perform as expected while SEXP is a good complement to the features selected. In selecting LT, PD and SEXP, our findings reinforce the assertion of \cite{Mockus2000} and \cite{Nagappan2005} while SEXP in the case of the work from \cite{Matsumoto2010} and \cite{Pascarella2019}.

In assessing LT and PD individually, would be zero valued if the change introduced new files. Considering that in new projects for software companies, introducing completely new files into the code repository is not uncommon. Since \cite{Kamei2013} used open source projects, the capture of the data may not represent newly created projects.

PD pertains to how old the last change was, and that the older the change means it is less likely to have a defect introduced \cite{Kamei2013}. As pointed out by Nagappan et al., looking at short burst of changes indicate a higher probability of defectiveness \cite{Nagappan2005}. However, this feature may not accurately describe new files introduced in the dataset and hence, this feature may have difficulty determining a new file's defectiveness. 

\begin{figure}[htbp]
\centerline{\includegraphics[width=0.5\textwidth]{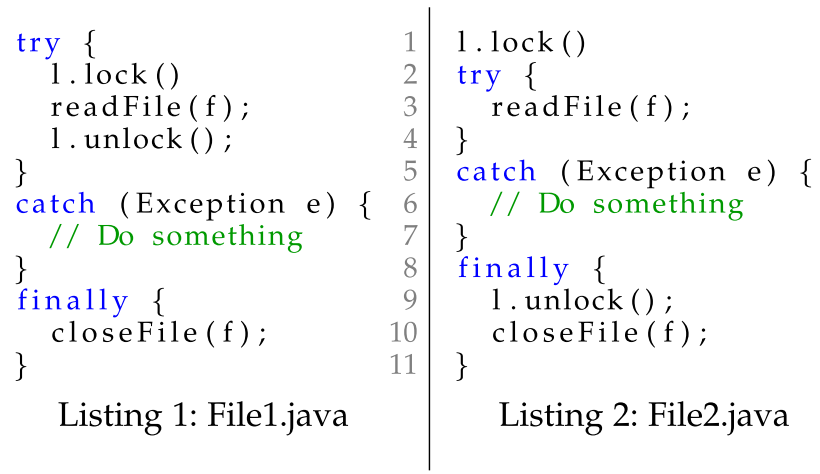}}
\caption{Code that could have similar metrics from \cite{Ghose2018}.}
\label{fig:ghose-similar-code}
\end{figure}

Using metrics to determine defectiveness can often overlook fine grained details. This was pointed out by \cite{Ghose2018}. The placement of a single line would have very similar code metrics. Figure \ref{fig:ghose-similar-code} illustrates such an issue.

Although the above shows a snippet of code, NF, NS, and ND would be the same where as other features could yield exactly the same calculated values.

The definition of software defects not only include problematic code but also code that does not conform to user specification \cite{Bourque2014,Ge2018}. Using that definition, context of code becomes important in determining defectiveness. We find that none of the features in the dataset establishes code context.

\section{Threats to Validity}
The manner in which the dataset was preprocessed ensured that the infinity values were avoided in the dataset. However, that also means that issues regarding the skewness of data would not be addressed in the features that were ignored. This may adversely affect the performances. However, we observed that the LT, and PD selection of features performed relatively stable across the different datasets. We also checked how the features affect the precision of defect prediction.  Table \ref{tab:precisiontable} shows the precision results 
of the features LT, PD, EXP and SEXP combinations with the Bugzilla dataset. The values shown are sorted in descending order.

\begin{table}[htbp]
\caption{Precision on Bugzilla Dataset on LT, PD, EXP, and SEXP}
\begin{center}
\begin{tabular}{ccc}
\toprule
\textbf{Features} & \textbf{Precision} \\
\midrule
LT,PD & 70.08\% \\
LT,PD,SEXP & 52.38\% \\
PD,SEXP & 51.85\% \\
LT,PD,EXP & 45.58 \% \\
PD,EXP & 43.62\% \\
\bottomrule
\end{tabular}
\label{tab:precisiontable}
\end{center}
\end{table}

As can be seen, LT,PD yielded a precision 70.08\% compared to LT,PD,SEXP at 52.38\%, PD,SEXP at 51.85\%, LT,PD,EXP at 45.58\%, and PD,EXP at 43.62\%. Hence, we believe the features will perform well on other datasets.

The considered datasets in the experiments were from Kamei's dataset and are Open-Source Systems (OSS). It is possible that commercial systems have different characteristics to open source ones as pointed out in \cite{Kamei2013}.

\section{Conclusion}
We found LT, and PD were identified as important features in training neural networks that predict defective code change submissions. These results were corroborated by the experiments that were conducted. However, in studying the features, we have come to identify that there needs to be features that can improve the determination of defective code. We also note that on newly contributed files into the source code repository, that the features present may not provide ample insight in determining defective code. We suggest that future works focus on doing more work in uncovering new features and adding more change-level datasets.

\section*{Acknowledgment}
We would like to thank the Software Technology department of De La Salle University, Dr. Charibeth Cheng, and Mr. Neil Patrick Del Gallego for their support and guidance in this research.

\bibliographystyle{ACM-Reference-Format}
\bibliography{biblio}


\begin{thebibliography}{00}


\ifx \showCODEN    \undefined \def \showCODEN     #1{\unskip}     \fi
\ifx \showDOI      \undefined \def \showDOI       #1{#1}\fi
\ifx \showISBNx    \undefined \def \showISBNx     #1{\unskip}     \fi
\ifx \showISBNxiii \undefined \def \showISBNxiii  #1{\unskip}     \fi
\ifx \showISSN     \undefined \def \showISSN      #1{\unskip}     \fi
\ifx \showLCCN     \undefined \def \showLCCN      #1{\unskip}     \fi
\ifx \shownote     \undefined \def \shownote      #1{#1}          \fi
\ifx \showarticletitle \undefined \def \showarticletitle #1{#1}   \fi
\ifx \showURL      \undefined \def \showURL       {\relax}        \fi
\providecommand\bibfield[2]{#2}
\providecommand\bibinfo[2]{#2}
\providecommand\natexlab[1]{#1}
\providecommand\showeprint[2][]{arXiv:#2}

\bibitem[\protect\citeauthoryear{Ablamowicz and Fauser}{Ablamowicz and
  Fauser}{2007}]%
        {Ablamowicz07}
\bibfield{author}{\bibinfo{person}{Rafal Ablamowicz} {and}
  \bibinfo{person}{Bertfried Fauser}.} \bibinfo{year}{2007}\natexlab{}.
\newblock \bibinfo{title}{CLIFFORD: a Maple 11 Package for Clifford Algebra
  Computations, version 11}.
\newblock   (\bibinfo{year}{2007}).
\newblock
\showURL{%
Retrieved February 28, 2008 from
  \url{http://math.tntech.edu/rafal/cliff11/index.html}}


\bibitem[\protect\citeauthoryear{Abril and Plant}{Abril and Plant}{2007}]%
        {Abril07}
\bibfield{author}{\bibinfo{person}{Patricia~S. Abril} {and}
  \bibinfo{person}{Robert Plant}.} \bibinfo{year}{2007}\natexlab{}.
\newblock \showarticletitle{The patent holder's dilemma: Buy, sell, or troll?}
\newblock \bibinfo{journal}{{\it Commun. ACM}} \bibinfo{volume}{50},
  \bibinfo{number}{1} (\bibinfo{date}{Jan.} \bibinfo{year}{2007}),
  \bibinfo{pages}{36--44}.
\newblock
\showDOI{%
\url{https://doi.org/10.1145/1188913.1188915}}


\bibitem[\protect\citeauthoryear{American Mathematical Society}{American
  Mathematical Society}{2015}]%
        {Amsthm15}
American Mathematical Society \bibinfo{year}{2015}\natexlab{}.
\newblock \bibinfo{booktitle}{{\em Using the amsthm Package}}.
\newblock American Mathematical Society.
\newblock
\newblock
\shownote{\url{http://www.ctan.org/pkg/amsthm}.}


\bibitem[\protect\citeauthoryear{Andler}{Andler}{1979}]%
        {Andler79}
\bibfield{author}{\bibinfo{person}{Sten Andler}.}
  \bibinfo{year}{1979}\natexlab{}.
\newblock \showarticletitle{Predicate Path expressions}. In
  \bibinfo{booktitle}{{\em Proceedings of the 6th. ACM SIGACT-SIGPLAN symposium
  on Principles of Programming Languages}} {\em (\bibinfo{series}{POPL '79})}.
  \bibinfo{publisher}{ACM Press}, \bibinfo{address}{New York, NY},
  \bibinfo{pages}{226--236}.
\newblock
\showDOI{%
\url{https://doi.org/10.1145/567752.567774}}


\bibitem[\protect\citeauthoryear{Anisi}{Anisi}{2003}]%
        {anisi03}
\bibfield{author}{\bibinfo{person}{David~A. Anisi}.}
  \bibinfo{year}{2003}\natexlab{}.
\newblock {\em \bibinfo{title}{Optimal Motion Control of a Ground Vehicle}}.
\newblock \bibinfo{thesistype}{Master's\ thesis}. \bibinfo{school}{Royal
  Institute of Technology (KTH), Stockholm, Sweden}.
\newblock


\bibitem[\protect\citeauthoryear{Bowman, Debray, and Peterson}{Bowman
  et~al\mbox{.}}{1993}]%
        {bowman:reasoning}
\bibfield{author}{\bibinfo{person}{Mic Bowman}, \bibinfo{person}{Saumya~K.
  Debray}, {and} \bibinfo{person}{Larry~L. Peterson}.}
  \bibinfo{year}{1993}\natexlab{}.
\newblock \showarticletitle{Reasoning About Naming Systems}.
\newblock \bibinfo{journal}{{\em ACM Trans. Program. Lang. Syst.\/}}
  \bibinfo{volume}{15}, \bibinfo{number}{5} (\bibinfo{date}{November}
  \bibinfo{year}{1993}), \bibinfo{pages}{795--825}.
\newblock
\showDOI{%
\url{https://doi.org/10.1145/161468.161471}}


\bibitem[\protect\citeauthoryear{Braams}{Braams}{1991}]%
        {braams:babel}
\bibfield{author}{\bibinfo{person}{Johannes Braams}.}
  \bibinfo{year}{1991}\natexlab{}.
\newblock \showarticletitle{Babel, a Multilingual Style-Option System for Use
  with LaTeX's Standard Document Styles}.
\newblock \bibinfo{journal}{{\em TUGboat\/}} \bibinfo{volume}{12},
  \bibinfo{number}{2} (\bibinfo{date}{June} \bibinfo{year}{1991}),
  \bibinfo{pages}{291--301}.
\newblock


\bibitem[\protect\citeauthoryear{Clark}{Clark}{1991}]%
        {clark:pct}
\bibfield{author}{\bibinfo{person}{Malcolm Clark}.}
  \bibinfo{year}{1991}\natexlab{}.
\newblock \showarticletitle{Post Congress Tristesse}. In
  \bibinfo{booktitle}{{\em TeX90 Conference Proceedings}}. TeX Users Group,
  \bibinfo{pages}{84--89}.
\newblock


\bibitem[\protect\citeauthoryear{Clarkson}{Clarkson}{1985}]%
        {Clarkson85}
\bibfield{author}{\bibinfo{person}{Kenneth~L. Clarkson}.}
  \bibinfo{year}{1985}\natexlab{}.
\newblock {\em \bibinfo{title}{Algorithms for Closest-Point Problems
  (Computational Geometry)}}.
\newblock \bibinfo{thesistype}{Ph.D. Dissertation}. \bibinfo{school}{Stanford
  University}, \bibinfo{address}{Palo Alto, CA}.
\newblock
\newblock
\shownote{UMI Order Number: AAT 8506171.}


\bibitem[\protect\citeauthoryear{Cohen}{Cohen}{1996}]%
        {JCohen96}
\bibfield{editor}{\bibinfo{person}{Jacques Cohen}} (Ed.).
  \bibinfo{year}{1996}\natexlab{}. \showarticletitle{Special issue: Digital
  Libraries}.
\newblock \bibinfo{journal}{{\em Commun. {ACM}\/}} \bibinfo{volume}{39},
  \bibinfo{number}{11} (\bibinfo{date}{Nov.} \bibinfo{year}{1996}).

\bibitem[\protect\citeauthoryear{Cohen, Nutt, and Sagic}{Cohen
  et~al\mbox{.}}{2007}]%
        {Cohen07}
\bibfield{author}{\bibinfo{person}{Sarah Cohen}, \bibinfo{person}{Werner Nutt},
  {and} \bibinfo{person}{Yehoshua Sagic}.} \bibinfo{year}{2007}\natexlab{}.
\newblock \showarticletitle{Deciding equivalances among conjunctive aggregate
  queries}.
\newblock \bibinfo{journal}{{\em J. ACM\/}} \bibinfo{volume}{54},
  \bibinfo{number}{2}, Article \bibinfo{articleno}{5} (\bibinfo{date}{April}
  \bibinfo{year}{2007}), \bibinfo{numpages}{50}~pages.
\newblock
\showDOI{%
\url{https://doi.org/10.1145/1219092.1219093}}


\bibitem[\protect\citeauthoryear{Douglass, Harel, and Trakhtenbrot}{Douglass
  et~al\mbox{.}}{1998}]%
        {Douglass98}
\bibfield{author}{\bibinfo{person}{Bruce~P. Douglass}, \bibinfo{person}{David
  Harel}, {and} \bibinfo{person}{Mark~B. Trakhtenbrot}.}
  \bibinfo{year}{1998}\natexlab{}.
\newblock \showarticletitle{Statecarts in use: structured analysis and
  object-orientation}.
\newblock In \bibinfo{booktitle}{{\em Lectures on Embedded Systems}},
  \bibfield{editor}{\bibinfo{person}{Grzegorz Rozenberg} {and}
  \bibinfo{person}{Frits~W. Vaandrager}} (Eds.). \bibinfo{series}{Lecture Notes
  in Computer Science}, Vol.~\bibinfo{volume}{1494}.
  \bibinfo{publisher}{Springer-Verlag}, \bibinfo{address}{London},
  \bibinfo{pages}{368--394}.
\newblock
\showDOI{%
\url{https://doi.org/10.1007/3-540-65193-4_29}}


\bibitem[\protect\citeauthoryear{Editor}{Editor}{2007}]%
        {Editor00}
\bibfield{editor}{\bibinfo{person}{Ian Editor}} (Ed.).
  \bibinfo{year}{2007}\natexlab{}.
\newblock \bibinfo{booktitle}{{\em The title of book one\/}
  (\bibinfo{edition}{1st.} ed.)}. \bibinfo{series}{The name of the series one},
  Vol.~\bibinfo{volume}{9}.
\newblock \bibinfo{publisher}{University of Chicago Press},
  \bibinfo{address}{Chicago}.
\newblock
\showDOI{%
\url{https://doi.org/10.1007/3-540-09237-4}}


\bibitem[\protect\citeauthoryear{Editor}{Editor}{2008}]%
        {Editor00a}
\bibfield{editor}{\bibinfo{person}{Ian Editor}} (Ed.).
  \bibinfo{year}{2008}\natexlab{}.
\newblock \bibinfo{booktitle}{{\em The title of book two\/}
  (\bibinfo{edition}{2nd.} ed.)}.
\newblock \bibinfo{publisher}{University of Chicago Press},
  \bibinfo{address}{Chicago}, Chapter 100.
\newblock
\showDOI{%
\url{https://doi.org/10.1007/3-540-09237-4}}


\bibitem[\protect\citeauthoryear{Fear}{Fear}{2005}]%
        {Fear05}
\bibfield{author}{\bibinfo{person}{Simon Fear}.}
  \bibinfo{year}{2005}\natexlab{}.
\newblock \bibinfo{booktitle}{{\em Publication quality tables in {\LaTeX}}}.
\newblock
\newblock
\shownote{\url{http://www.ctan.org/pkg/booktabs}.}


\bibitem[\protect\citeauthoryear{Gundy, Balzarotti, and Vigna}{Gundy
  et~al\mbox{.}}{2007}]%
        {VanGundy07}
\bibfield{author}{\bibinfo{person}{Matthew~Van Gundy}, \bibinfo{person}{Davide
  Balzarotti}, {and} \bibinfo{person}{Giovanni Vigna}.}
  \bibinfo{year}{2007}\natexlab{}.
\newblock \showarticletitle{Catch me, if you can: Evading network signatures
  with web-based polymorphic worms}. In \bibinfo{booktitle}{{\em Proceedings of
  the first USENIX workshop on Offensive Technologies}} {\em
  (\bibinfo{series}{WOOT '07})}. \bibinfo{publisher}{USENIX Association},
  \bibinfo{address}{Berkley, CA}, Article \bibinfo{articleno}{7},
  \bibinfo{numpages}{9}~pages.
\newblock


\bibitem[\protect\citeauthoryear{Harel}{Harel}{1978}]%
        {Harel78}
\bibfield{author}{\bibinfo{person}{David Harel}.}
  \bibinfo{year}{1978}\natexlab{}.
\newblock \bibinfo{booktitle}{{\em LOGICS of Programs: AXIOMATICS and
  DESCRIPTIVE POWER}}.
\newblock \bibinfo{type}{MIT Research Lab Technical Report} TR-200.
  \bibinfo{institution}{Massachusetts Institute of Technology},
  \bibinfo{address}{Cambridge, MA}.
\newblock


\bibitem[\protect\citeauthoryear{Harel}{Harel}{1979}]%
        {Harel79}
\bibfield{author}{\bibinfo{person}{David Harel}.}
  \bibinfo{year}{1979}\natexlab{}.
\newblock \bibinfo{booktitle}{{\em First-Order Dynamic Logic}}.
  \bibinfo{series}{Lecture Notes in Computer Science},
  Vol.~\bibinfo{volume}{68}.
\newblock \bibinfo{publisher}{Springer-Verlag}, \bibinfo{address}{New York,
  NY}.
\newblock
\showDOI{%
\url{https://doi.org/10.1007/3-540-09237-4}}


\bibitem[\protect\citeauthoryear{Herlihy}{Herlihy}{1993}]%
        {herlihy:methodology}
\bibfield{author}{\bibinfo{person}{Maurice Herlihy}.}
  \bibinfo{year}{1993}\natexlab{}.
\newblock \showarticletitle{A Methodology for Implementing Highly Concurrent
  Data Objects}.
\newblock \bibinfo{journal}{{\em ACM Trans. Program. Lang. Syst.\/}}
  \bibinfo{volume}{15}, \bibinfo{number}{5} (\bibinfo{date}{November}
  \bibinfo{year}{1993}), \bibinfo{pages}{745--770}.
\newblock
\showDOI{%
\url{https://doi.org/10.1145/161468.161469}}


\bibitem[\protect\citeauthoryear{H{\"o}rmander}{H{\"o}rmander}{1985a}]%
        {MR781537}
\bibfield{author}{\bibinfo{person}{Lars H{\"o}rmander}.}
  \bibinfo{year}{1985}\natexlab{a}.
\newblock \bibinfo{booktitle}{{\em The analysis of linear partial differential
  operators. {III}}}. \bibinfo{series}{Grundlehren der Mathematischen
  Wissenschaften [Fundamental Principles of Mathematical Sciences]},
  Vol.~\bibinfo{volume}{275}.
\newblock \bibinfo{publisher}{Springer-Verlag}, \bibinfo{address}{Berlin,
  Germany}. viii+525 pages.
\newblock
\showISBNx{3-540-13828-5}
\newblock
\shownote{Pseudodifferential operators.}


\bibitem[\protect\citeauthoryear{H{\"o}rmander}{H{\"o}rmander}{1985b}]%
        {MR781536}
\bibfield{author}{\bibinfo{person}{Lars H{\"o}rmander}.}
  \bibinfo{year}{1985}\natexlab{b}.
\newblock \bibinfo{booktitle}{{\em The analysis of linear partial differential
  operators. {IV}}}. \bibinfo{series}{Grundlehren der Mathematischen
  Wissenschaften [Fundamental Principles of Mathematical Sciences]},
  Vol.~\bibinfo{volume}{275}.
\newblock \bibinfo{publisher}{Springer-Verlag}, \bibinfo{address}{Berlin,
  Germany}. vii+352 pages.
\newblock
\showISBNx{3-540-13829-3}
\newblock
\shownote{Fourier integral operators.}


\bibitem[\protect\citeauthoryear{IEEE}{IEEE}{2004}]%
        {2004:ITE:1009386.1010128}
IEEE \bibinfo{year}{2004}\natexlab{}.
\newblock \showarticletitle{IEEE TCSC Executive Committee}. In
  \bibinfo{booktitle}{{\em Proceedings of the IEEE International Conference on
  Web Services}} {\em (\bibinfo{series}{ICWS '04})}. \bibinfo{publisher}{IEEE
  Computer Society}, \bibinfo{address}{Washington, DC, USA},
  \bibinfo{pages}{21--22}.
\newblock
\showISBNx{0-7695-2167-3}
\showDOI{%
\url{https://doi.org/10.1109/ICWS.2004.64}}


\bibitem[\protect\citeauthoryear{Kirschmer and Voight}{Kirschmer and
  Voight}{2010}]%
        {Kirschmer:2010:AEI:1958016.1958018}
\bibfield{author}{\bibinfo{person}{Markus Kirschmer} {and}
  \bibinfo{person}{John Voight}.} \bibinfo{year}{2010}\natexlab{}.
\newblock \showarticletitle{Algorithmic Enumeration of Ideal Classes for
  Quaternion Orders}.
\newblock \bibinfo{journal}{{\em SIAM J. Comput.\/}} \bibinfo{volume}{39},
  \bibinfo{number}{5} (\bibinfo{date}{Jan.} \bibinfo{year}{2010}),
  \bibinfo{pages}{1714--1747}.
\newblock
\showISSN{0097-5397}
\showDOI{%
\url{https://doi.org/10.1137/080734467}}


\bibitem[\protect\citeauthoryear{Knuth}{Knuth}{1997}]%
        {Knuth97}
\bibfield{author}{\bibinfo{person}{Donald~E. Knuth}.}
  \bibinfo{year}{1997}\natexlab{}.
\newblock \bibinfo{booktitle}{{\em The Art of Computer Programming, Vol. 1:
  Fundamental Algorithms (3rd. ed.)}}.
\newblock \bibinfo{publisher}{Addison Wesley Longman Publishing Co., Inc.}
\newblock


\bibitem[\protect\citeauthoryear{Kosiur}{Kosiur}{2001}]%
        {Kosiur01}
\bibfield{author}{\bibinfo{person}{David Kosiur}.}
  \bibinfo{year}{2001}\natexlab{}.
\newblock \bibinfo{booktitle}{{\em Understanding Policy-Based Networking\/}
  (\bibinfo{edition}{2nd.} ed.)}.
\newblock \bibinfo{publisher}{Wiley}, \bibinfo{address}{New York, NY}.
\newblock


\bibitem[\protect\citeauthoryear{Lamport}{Lamport}{1986}]%
        {Lamport:LaTeX}
\bibfield{author}{\bibinfo{person}{Leslie Lamport}.}
  \bibinfo{year}{1986}\natexlab{}.
\newblock \bibinfo{booktitle}{{\em \it {\LaTeX}: A Document Preparation
  System}}.
\newblock \bibinfo{publisher}{Addison-Wesley}, \bibinfo{address}{Reading, MA.}
\newblock


\bibitem[\protect\citeauthoryear{Lee}{Lee}{2005}]%
        {Lee05}
\bibfield{author}{\bibinfo{person}{Newton Lee}.}
  \bibinfo{year}{2005}\natexlab{}.
\newblock \showarticletitle{Interview with Bill Kinder: January 13, 2005}.
\newblock \bibinfo{howpublished}{Video}.
\newblock \bibinfo{journal}{{\em Comput. Entertain.\/}} \bibinfo{volume}{3},
  \bibinfo{number}{1}, Article \bibinfo{articleno}{4}
  (\bibinfo{date}{Jan.-March} \bibinfo{year}{2005}).
\newblock
\showDOI{%
\url{https://doi.org/10.1145/1057270.1057278}}


\bibitem[\protect\citeauthoryear{Novak}{Novak}{2003}]%
        {Novak03}
\bibfield{author}{\bibinfo{person}{Dave Novak}.}
  \bibinfo{year}{2003}\natexlab{}.
\newblock \showarticletitle{Solder man}. \bibinfo{howpublished}{Video}. In
  \bibinfo{booktitle}{{\em ACM SIGGRAPH 2003 Video Review on Animation theater
  Program: Part I - Vol. 145 (July 27--27, 2003)}}. \bibinfo{publisher}{ACM
  Press}, \bibinfo{address}{New York, NY}, \bibinfo{pages}{4}.
\newblock
\showDOI{%
\url{https://doi.org/99.9999/woot07-S422}}


\bibitem[\protect\citeauthoryear{Obama}{Obama}{2008}]%
        {Obama08}
\bibfield{author}{\bibinfo{person}{Barack Obama}.}
  \bibinfo{year}{2008}\natexlab{}.
\newblock \bibinfo{title}{A more perfect union}.
\newblock \bibinfo{howpublished}{Video}.   (\bibinfo{date}{5 March}
  \bibinfo{year}{2008}).
\newblock
\showURL{%
Retrieved March 21, 2008 from
  \url{http://video.google.com/videoplay?docid=6528042696351994555}}


\bibitem[\protect\citeauthoryear{Poker-Edge.Com}{Poker-Edge.Com}{2006}]%
        {Poker06}
\bibfield{author}{\bibinfo{person}{Poker-Edge.Com}.}
  \bibinfo{year}{2006}\natexlab{}.
\newblock \bibinfo{title}{Stats and Analysis}.
\newblock   (\bibinfo{date}{March} \bibinfo{year}{2006}).
\newblock
\showURL{%
Retrieved June 7, 2006 from \url{http://www.poker-edge.com/stats.php}}


\bibitem[\protect\citeauthoryear{Rous}{Rous}{2008}]%
        {rous08}
\bibfield{author}{\bibinfo{person}{Bernard Rous}.}
  \bibinfo{year}{2008}\natexlab{}.
\newblock \showarticletitle{The Enabling of Digital Libraries}.
\newblock \bibinfo{journal}{{\em Digital Libraries\/}} \bibinfo{volume}{12},
  \bibinfo{number}{3}, Article \bibinfo{articleno}{5} (\bibinfo{date}{July}
  \bibinfo{year}{2008}).
\newblock
\newblock
\shownote{To appear.}


\bibitem[\protect\citeauthoryear{Saeedi, Zamani, and Sedighi}{Saeedi
  et~al\mbox{.}}{2010a}]%
        {SaeediMEJ10}
\bibfield{author}{\bibinfo{person}{Mehdi Saeedi},
  \bibinfo{person}{Morteza~Saheb Zamani}, {and} \bibinfo{person}{Mehdi
  Sedighi}.} \bibinfo{year}{2010}\natexlab{a}.
\newblock \showarticletitle{A library-based synthesis methodology for
  reversible logic}.
\newblock \bibinfo{journal}{{\em Microelectron. J.\/}} \bibinfo{volume}{41},
  \bibinfo{number}{4} (\bibinfo{date}{April} \bibinfo{year}{2010}),
  \bibinfo{pages}{185--194}.
\newblock


\bibitem[\protect\citeauthoryear{Saeedi, Zamani, Sedighi, and Sasanian}{Saeedi
  et~al\mbox{.}}{2010b}]%
        {SaeediJETC10}
\bibfield{author}{\bibinfo{person}{Mehdi Saeedi},
  \bibinfo{person}{Morteza~Saheb Zamani}, \bibinfo{person}{Mehdi Sedighi},
  {and} \bibinfo{person}{Zahra Sasanian}.} \bibinfo{year}{2010}\natexlab{b}.
\newblock \showarticletitle{Synthesis of Reversible Circuit Using Cycle-Based
  Approach}.
\newblock \bibinfo{journal}{{\em J. Emerg. Technol. Comput. Syst.\/}}
  \bibinfo{volume}{6}, \bibinfo{number}{4} (\bibinfo{date}{Dec.}
  \bibinfo{year}{2010}).
\newblock


\bibitem[\protect\citeauthoryear{Salas and Hille}{Salas and Hille}{1978}]%
        {salas:calculus}
\bibfield{author}{\bibinfo{person}{S.L. Salas} {and} \bibinfo{person}{Einar
  Hille}.} \bibinfo{year}{1978}\natexlab{}.
\newblock \bibinfo{booktitle}{{\em Calculus: One and Several Variable}}.
\newblock \bibinfo{publisher}{John Wiley and Sons}, \bibinfo{address}{New
  York}.
\newblock


\bibitem[\protect\citeauthoryear{Scientist}{Scientist}{2009}]%
        {JoeScientist001}
\bibfield{author}{\bibinfo{person}{Joseph Scientist}.}
  \bibinfo{year}{2009}\natexlab{}.
\newblock \bibinfo{title}{The fountain of youth}.
\newblock   (\bibinfo{date}{Aug.} \bibinfo{year}{2009}).
\newblock
\newblock
\shownote{Patent No. 12345, Filed July 1st., 2008, Issued Aug. 9th., 2009.}


\bibitem[\protect\citeauthoryear{Smith}{Smith}{2010}]%
        {Smith10}
\bibfield{author}{\bibinfo{person}{Stan~W. Smith}.}
  \bibinfo{year}{2010}\natexlab{}.
\newblock \showarticletitle{An experiment in bibliographic mark-up: Parsing
  metadata for XML export}. In \bibinfo{booktitle}{{\em Proceedings of the 3rd.
  annual workshop on Librarians and Computers}} {\em (\bibinfo{series}{LAC
  '10})}, \bibfield{editor}{\bibinfo{person}{Reginald~N. Smythe} {and}
  \bibinfo{person}{Alexander Noble}} (Eds.), Vol.~\bibinfo{volume}{3}.
  \bibinfo{publisher}{Paparazzi Press}, \bibinfo{address}{Milan Italy},
  \bibinfo{pages}{422--431}.
\newblock
\showDOI{%
\url{https://doi.org/99.9999/woot07-S422}}


\bibitem[\protect\citeauthoryear{Spector}{Spector}{1990}]%
        {Spector90}
\bibfield{author}{\bibinfo{person}{Asad~Z. Spector}.}
  \bibinfo{year}{1990}\natexlab{}.
\newblock \showarticletitle{Achieving application requirements}.
\newblock In \bibinfo{booktitle}{{\em Distributed Systems}
  (\bibinfo{edition}{2nd.} ed.)}, \bibfield{editor}{\bibinfo{person}{Sape
  Mullender}} (Ed.). \bibinfo{publisher}{ACM Press}, \bibinfo{address}{New
  York, NY}, \bibinfo{pages}{19--33}.
\newblock
\showDOI{%
\url{https://doi.org/10.1145/90417.90738}}


\bibitem[\protect\citeauthoryear{Thornburg}{Thornburg}{2001}]%
        {Thornburg01}
\bibfield{author}{\bibinfo{person}{Harry Thornburg}.}
  \bibinfo{year}{2001}\natexlab{}.
\newblock \bibinfo{title}{Introduction to Bayesian Statistics}.
\newblock   (\bibinfo{date}{March} \bibinfo{year}{2001}).
\newblock
\showURL{%
Retrieved March 2, 2005 from
  \url{http://ccrma.stanford.edu/~jos/bayes/bayes.html}}


\bibitem[\protect\citeauthoryear{TUG}{TUG}{2017}]%
        {TUGInstmem}
TUG \bibinfo{year}{2017}\natexlab{}.
\newblock \bibinfo{title}{Institutional members of the {\TeX} Users Group}.
\newblock   (\bibinfo{year}{2017}).
\newblock
\showURL{%
Retrieved May 27, 2017 from \url{http://wwtug.org/instmem.html}}


\bibitem[\protect\citeauthoryear{Veytsman}{Veytsman}{}]%
        {CTANacmart}
\bibfield{author}{\bibinfo{person}{Boris Veytsman}.}
\newblock \bibinfo{title}{acmart---{C}lass for typesetting publications of
  {ACM}}.
\newblock   (\bibinfo{year}{????}).
\newblock
\showURL{%
Retrieved May 27, 2017 from \url{http://www.ctan.org/pkg/acmart}}


\end{thebibliography}



\begin{thebibliography}{31}


\ifx \showCODEN    \undefined \def \showCODEN     #1{\unskip}     \fi
\ifx \showDOI      \undefined \def \showDOI       #1{#1}\fi
\ifx \showISBNx    \undefined \def \showISBNx     #1{\unskip}     \fi
\ifx \showISBNxiii \undefined \def \showISBNxiii  #1{\unskip}     \fi
\ifx \showISSN     \undefined \def \showISSN      #1{\unskip}     \fi
\ifx \showLCCN     \undefined \def \showLCCN      #1{\unskip}     \fi
\ifx \shownote     \undefined \def \shownote      #1{#1}          \fi
\ifx \showarticletitle \undefined \def \showarticletitle #1{#1}   \fi
\ifx \showURL      \undefined \def \showURL       {\relax}        \fi
\providecommand\bibfield[2]{#2}
\providecommand\bibinfo[2]{#2}
\providecommand\natexlab[1]{#1}
\providecommand\showeprint[2][]{arXiv:#2}

\bibitem[\protect\citeauthoryear{Alsaeedi and Khan}{Alsaeedi and Khan}{2019}]%
        {Alsaeedi2019}
\bibfield{author}{\bibinfo{person}{Abdullah Alsaeedi} {and}
  \bibinfo{person}{Mohammad~Zubair Khan}.} \bibinfo{year}{2019}\natexlab{}.
\newblock \showarticletitle{{Software Defect Prediction Using Supervised
  Machine Learning and Ensemble Techniques: A Comparative Study}}.
\newblock \bibinfo{journal}{\emph{Journal of Software Engineering and
  Applications}} \bibinfo{volume}{12}, \bibinfo{number}{05}
  (\bibinfo{year}{2019}), \bibinfo{pages}{85--100}.
\newblock
\showISSN{1945-3116}
\urldef\tempurl%
\url{https://doi.org/10.4236/jsea.2019.125007}
\showDOI{\tempurl}


\bibitem[\protect\citeauthoryear{Bellego and Pape}{Bellego and Pape}{2019}]%
        {Bellego2019}
\bibfield{author}{\bibinfo{person}{Christophe Bellego} {and}
  \bibinfo{person}{Louis-Daniel Pape}.} \bibinfo{year}{2019}\natexlab{}.
\newblock \showarticletitle{{Dealing with the log of zero in regression
  models}}.
\newblock \bibinfo{journal}{\emph{S{\'{e}}rie des Documents de Travail}}
  (\bibinfo{year}{2019}), \bibinfo{pages}{16}.
\newblock
\showISSN{1556-5068}
\urldef\tempurl%
\url{https://doi.org/10.2139/ssrn.3444996}
\showDOI{\tempurl}


\bibitem[\protect\citeauthoryear{Bennin, Keung, Phannachitta, Monden, and
  Mensah}{Bennin et~al\mbox{.}}{2018}]%
        {Bennin2018}
\bibfield{author}{\bibinfo{person}{Kwabena~Ebo Bennin}, \bibinfo{person}{Jacky
  Keung}, \bibinfo{person}{Passakorn Phannachitta}, \bibinfo{person}{Akito
  Monden}, {and} \bibinfo{person}{Solomon Mensah}.}
  \bibinfo{year}{2018}\natexlab{}.
\newblock \showarticletitle{{MAHAKIL: Diversity Based Oversampling Approach to
  Alleviate the Class Imbalance Issue in Software Defect Prediction}}.
\newblock \bibinfo{journal}{\emph{IEEE Transactions on Software Engineering}}
  \bibinfo{volume}{44}, \bibinfo{number}{6} (\bibinfo{year}{2018}),
  \bibinfo{pages}{534--550}.
\newblock
\showISSN{00985589}
\urldef\tempurl%
\url{https://doi.org/10.1109/TSE.2017.2731766}
\showDOI{\tempurl}


\bibitem[\protect\citeauthoryear{Bo and Li}{Bo and Li}{2008}]%
        {Bo2008}
\bibfield{author}{\bibinfo{person}{Rui Bo} {and} \bibinfo{person}{Fangxing
  Li}.} \bibinfo{year}{2008}\natexlab{}.
\newblock \showarticletitle{{Power flow studies using Principal Component
  Analysis}}.
\newblock \bibinfo{journal}{\emph{40th North American Power Symposium,
  NAPS2008}} \bibinfo{number}{1} (\bibinfo{year}{2008}), \bibinfo{pages}{1--6}.
\newblock
\showISBNx{9781424442836}
\urldef\tempurl%
\url{https://doi.org/10.1109/NAPS.2008.5307323}
\showDOI{\tempurl}


\bibitem[\protect\citeauthoryear{Borg, Svensson, Berg, and Hansson}{Borg
  et~al\mbox{.}}{2019}]%
        {Borg2019}
\bibfield{author}{\bibinfo{person}{Markus Borg}, \bibinfo{person}{Oscar
  Svensson}, \bibinfo{person}{Kristian Berg}, {and} \bibinfo{person}{Daniel
  Hansson}.} \bibinfo{year}{2019}\natexlab{}.
\newblock \showarticletitle{{SZZ unleashed: an open implementation of the SZZ
  algorithm - featuring example usage in a study of just-in-time bug prediction
  for the Jenkins project}}. In \bibinfo{booktitle}{\emph{Proceedings of the
  3rd ACM SIGSOFT International Workshop on Machine Learning Techniques for
  Software Quality Evaluation - MaLTeSQuE 2019}}. \bibinfo{publisher}{ACM
  Press}, \bibinfo{address}{New York, New York, USA}, \bibinfo{pages}{7--12}.
\newblock
\showISBNx{9781450368551}
\urldef\tempurl%
\url{https://doi.org/10.1145/3340482.3342742}
\showDOI{\tempurl}
\showeprint[arxiv]{1903.01742}


\bibitem[\protect\citeauthoryear{Bourque and Fairley}{Bourque and
  Fairley}{2014}]%
        {Bourque2014}
\bibfield{author}{\bibinfo{person}{Pierre Bourque} {and}
  \bibinfo{person}{Richard~E. Fairley}.} \bibinfo{year}{2014}\natexlab{}.
\newblock \bibinfo{booktitle}{\emph{{SWEBOK v.3 - Guide to the Software
  Engineering - Body of Knowledge.}}}
\newblock 346 pages.
\newblock
\showISBNx{0-7695-2330-7}
\showISSN{07407459}
\urldef\tempurl%
\url{https://doi.org/10.1234/12345678}
\showDOI{\tempurl}


\bibitem[\protect\citeauthoryear{Cai, Fan, Yan, and Xia}{Cai
  et~al\mbox{.}}{2019}]%
        {82ac60c1c56b4be4a6220de0dfad6039}
\bibfield{author}{\bibinfo{person}{Liang Cai}, \bibinfo{person}{Yuan~Rui Fan},
  \bibinfo{person}{Meng Yan}, {and} \bibinfo{person}{Xin Xia}.}
  \bibinfo{year}{2019}\natexlab{}.
\newblock \showarticletitle{{Just-in-time software defect prediction:
  literature review}}.
\newblock \bibinfo{journal}{\emph{Ruan Jian Xue Bao/Journal of Software}}
  \bibinfo{volume}{30}, \bibinfo{number}{5} (\bibinfo{year}{2019}),
  \bibinfo{pages}{1288--1307}.
\newblock
\showISSN{1000-9825}
\urldef\tempurl%
\url{https://doi.org/10.13328/j.cnki.jos.005713}
\showDOI{\tempurl}


\bibitem[\protect\citeauthoryear{D'Ambros, Lanza, and Robbes}{D'Ambros
  et~al\mbox{.}}{2010}]%
        {DAmbros2010}
\bibfield{author}{\bibinfo{person}{Marco D'Ambros}, \bibinfo{person}{Michele
  Lanza}, {and} \bibinfo{person}{Romain Robbes}.}
  \bibinfo{year}{2010}\natexlab{}.
\newblock \showarticletitle{{An extensive comparison of bug prediction
  approaches}}.
\newblock \bibinfo{journal}{\emph{Proceedings - International Conference on
  Software Engineering}} \bibinfo{number}{May 2010} (\bibinfo{year}{2010}),
  \bibinfo{pages}{31--41}.
\newblock
\showISBNx{9781424468034}
\showISSN{02705257}
\urldef\tempurl%
\url{https://doi.org/10.1109/MSR.2010.5463279}
\showDOI{\tempurl}


\bibitem[\protect\citeauthoryear{Fu and Menzies}{Fu and Menzies}{2017}]%
        {Fu2017}
\bibfield{author}{\bibinfo{person}{Wei Fu} {and} \bibinfo{person}{Tim
  Menzies}.} \bibinfo{year}{2017}\natexlab{}.
\newblock \showarticletitle{{Revisiting unsupervised learning for defect
  prediction}}.
\newblock  (\bibinfo{year}{2017}), \bibinfo{pages}{72--83}.
\newblock
\showISBNx{9781450351058}
\urldef\tempurl%
\url{https://doi.org/10.1145/3106237.3106257}
\showDOI{\tempurl}
\showeprint[arxiv]{1703.00132}


\bibitem[\protect\citeauthoryear{Ge, Liu, and Liu}{Ge et~al\mbox{.}}{2018}]%
        {Ge2018}
\bibfield{author}{\bibinfo{person}{Jianxin Ge}, \bibinfo{person}{Jiaomin Liu},
  {and} \bibinfo{person}{Wenyuan Liu}.} \bibinfo{year}{2018}\natexlab{}.
\newblock \showarticletitle{{Comparative study on defect prediction algorithms
  of supervised learning software based on imbalanced classification data
  sets}}.
\newblock \bibinfo{journal}{\emph{Proceedings - 2018 IEEE/ACIS 19th
  International Conference on Software Engineering, Artificial Intelligence,
  Networking and Parallel/Distributed Computing, SNPD 2018}}
  (\bibinfo{year}{2018}), \bibinfo{pages}{399--406}.
\newblock
\showISBNx{9781538658895}
\urldef\tempurl%
\url{https://doi.org/10.1109/SNPD.2018.8441143}
\showDOI{\tempurl}


\bibitem[\protect\citeauthoryear{Ghose, Ng, Tran, Pham, Grundy, and Dam}{Ghose
  et~al\mbox{.}}{2018}]%
        {Ghose2018}
\bibfield{author}{\bibinfo{person}{Aditya Ghose}, \bibinfo{person}{Shien~Wee
  Ng}, \bibinfo{person}{Truyen Tran}, \bibinfo{person}{Trang Thi~Minh Pham},
  \bibinfo{person}{John Grundy}, {and} \bibinfo{person}{Hoa~Khanh Dam}.}
  \bibinfo{year}{2018}\natexlab{}.
\newblock \showarticletitle{{Automatic feature learning for predicting
  vulnerable software components}}.
\newblock \bibinfo{journal}{\emph{IEEE Transactions on Software Engineering}}
  \bibinfo{volume}{14}, \bibinfo{number}{November} (\bibinfo{year}{2018}),
  \bibinfo{pages}{1--1}.
\newblock
\showISSN{0098-5589}
\urldef\tempurl%
\url{https://doi.org/10.1109/tse.2018.2881961}
\showDOI{\tempurl}


\bibitem[\protect\citeauthoryear{Hassan}{Hassan}{2009}]%
        {Hassan2009}
\bibfield{author}{\bibinfo{person}{Ahmed~E. Hassan}.}
  \bibinfo{year}{2009}\natexlab{}.
\newblock \showarticletitle{{Predicting faults using the complexity of code
  changes}}.
\newblock \bibinfo{journal}{\emph{Proceedings - International Conference on
  Software Engineering}} (\bibinfo{year}{2009}), \bibinfo{pages}{78--88}.
\newblock
\showISBNx{9781424434527}
\showISSN{02705257}
\urldef\tempurl%
\url{https://doi.org/10.1109/ICSE.2009.5070510}
\showDOI{\tempurl}


\bibitem[\protect\citeauthoryear{Hosseini, Turhan, and Gunarathna}{Hosseini
  et~al\mbox{.}}{2019}]%
        {Hosseini2019}
\bibfield{author}{\bibinfo{person}{Seyedrebvar Hosseini},
  \bibinfo{person}{Burak Turhan}, {and} \bibinfo{person}{Dimuthu Gunarathna}.}
  \bibinfo{year}{2019}\natexlab{}.
\newblock \showarticletitle{{A systematic literature review and meta-analysis
  on cross project defect prediction}}.
\newblock \bibinfo{journal}{\emph{IEEE Transactions on Software Engineering}}
  \bibinfo{volume}{45}, \bibinfo{number}{2} (\bibinfo{year}{2019}),
  \bibinfo{pages}{111--147}.
\newblock
\showISSN{19393520}
\urldef\tempurl%
\url{https://doi.org/10.1109/TSE.2017.2770124}
\showDOI{\tempurl}


\bibitem[\protect\citeauthoryear{Huang, Xia, and Lo}{Huang
  et~al\mbox{.}}{2017}]%
        {Huang2017}
\bibfield{author}{\bibinfo{person}{Qiao Huang}, \bibinfo{person}{Xin Xia},
  {and} \bibinfo{person}{David Lo}.} \bibinfo{year}{2017}\natexlab{}.
\newblock \showarticletitle{{Supervised vs Unsupervised Models: A Holistic Look
  at Effort-Aware Just-in-Time Defect Prediction}}. In
  \bibinfo{booktitle}{\emph{2017 IEEE International Conference on Software
  Maintenance and Evolution (ICSME)}}. \bibinfo{publisher}{IEEE},
  \bibinfo{pages}{159--170}.
\newblock
\showISBNx{978-1-5386-0992-7}
\urldef\tempurl%
\url{https://doi.org/10.1109/ICSME.2017.51}
\showDOI{\tempurl}


\bibitem[\protect\citeauthoryear{Huang, Xia, and Lo}{Huang
  et~al\mbox{.}}{2018}]%
        {Huang2018}
\bibfield{author}{\bibinfo{person}{Qiao Huang}, \bibinfo{person}{Xin Xia},
  {and} \bibinfo{person}{David Lo}.} \bibinfo{year}{2018}\natexlab{}.
\newblock \bibinfo{booktitle}{\emph{{Revisiting supervised and unsupervised
  models for effort-aware just-in-time defect prediction}}}.
\newblock \bibinfo{publisher}{Empirical Software Engineering}.
\newblock
\showISBNx{1066401896}
\showISSN{15737616}
\urldef\tempurl%
\url{https://doi.org/10.1007/s10664-018-9661-2}
\showDOI{\tempurl}


\bibitem[\protect\citeauthoryear{Kamei, Shihab, Adams, Hassan, Mockus, Sinha,
  and Ubayashi}{Kamei et~al\mbox{.}}{2013}]%
        {Kamei2013}
\bibfield{author}{\bibinfo{person}{Yasutaka Kamei}, \bibinfo{person}{Emad
  Shihab}, \bibinfo{person}{Bram Adams}, \bibinfo{person}{Ahmed~E. Hassan},
  \bibinfo{person}{Audris Mockus}, \bibinfo{person}{Anand Sinha}, {and}
  \bibinfo{person}{Naoyasu Ubayashi}.} \bibinfo{year}{2013}\natexlab{}.
\newblock \showarticletitle{{A large-scale empirical study of just-in-time
  quality assurance}}.
\newblock \bibinfo{journal}{\emph{IEEE Transactions on Software Engineering}}
  \bibinfo{volume}{39}, \bibinfo{number}{6} (\bibinfo{year}{2013}),
  \bibinfo{pages}{757--773}.
\newblock
\showISSN{00985589}
\urldef\tempurl%
\url{https://doi.org/10.1109/TSE.2012.70}
\showDOI{\tempurl}


\bibitem[\protect\citeauthoryear{Lever, Krzywinski, and Altman}{Lever
  et~al\mbox{.}}{2017}]%
        {Lever2017}
\bibfield{author}{\bibinfo{person}{Jake Lever}, \bibinfo{person}{Martin
  Krzywinski}, {and} \bibinfo{person}{Naomi Altman}.}
  \bibinfo{year}{2017}\natexlab{}.
\newblock \showarticletitle{{Points of Significance: Principal component
  analysis}}.
\newblock \bibinfo{journal}{\emph{Nature Methods}} \bibinfo{volume}{14},
  \bibinfo{number}{7} (\bibinfo{year}{2017}), \bibinfo{pages}{641--642}.
\newblock
\showISSN{15487105}
\urldef\tempurl%
\url{https://doi.org/10.1038/nmeth.4346}
\showDOI{\tempurl}


\bibitem[\protect\citeauthoryear{Li, Jing, and Zhu}{Li et~al\mbox{.}}{2018}]%
        {Li2018}
\bibfield{author}{\bibinfo{person}{Zhiqiang Li}, \bibinfo{person}{Xiao-Yuan
  Jing}, {and} \bibinfo{person}{Xiaoke Zhu}.} \bibinfo{year}{2018}\natexlab{}.
\newblock \showarticletitle{{Progress on approaches to software defect
  prediction}}.
\newblock \bibinfo{journal}{\emph{IET Software}} \bibinfo{volume}{12},
  \bibinfo{number}{3} (\bibinfo{year}{2018}), \bibinfo{pages}{161--175}.
\newblock
\showISSN{1751-8806}
\urldef\tempurl%
\url{https://doi.org/10.1049/iet-sen.2017.0148}
\showDOI{\tempurl}


\bibitem[\protect\citeauthoryear{Matsumoto, Kamei, Monden, Matsumoto, and
  Nakamura}{Matsumoto et~al\mbox{.}}{2010}]%
        {Matsumoto2010}
\bibfield{author}{\bibinfo{person}{Shinsuke Matsumoto},
  \bibinfo{person}{Yasutaka Kamei}, \bibinfo{person}{Akito Monden},
  \bibinfo{person}{Ken~Ichi Matsumoto}, {and} \bibinfo{person}{Masahide
  Nakamura}.} \bibinfo{year}{2010}\natexlab{}.
\newblock \showarticletitle{{An analysis of developer metrics for fault
  prediction}}.
\newblock \bibinfo{journal}{\emph{ACM International Conference Proceeding
  Series}} \bibinfo{number}{January} (\bibinfo{year}{2010}).
\newblock
\showISBNx{9781450304047}
\urldef\tempurl%
\url{https://doi.org/10.1145/1868328.1868356}
\showDOI{\tempurl}


\bibitem[\protect\citeauthoryear{Mockus and Weiss}{Mockus and Weiss}{2000}]%
        {Mockus2000}
\bibfield{author}{\bibinfo{person}{Audris Mockus} {and}
  \bibinfo{person}{David~M. Weiss}.} \bibinfo{year}{2000}\natexlab{}.
\newblock \showarticletitle{{Predicting risk of software changes}}.
\newblock \bibinfo{journal}{\emph{Bell Labs Technical Journal}}
  \bibinfo{volume}{5}, \bibinfo{number}{2} (\bibinfo{year}{2000}),
  \bibinfo{pages}{169--180}.
\newblock
\showISSN{10897089}
\urldef\tempurl%
\url{https://doi.org/10.1002/bltj.2229}
\showDOI{\tempurl}


\bibitem[\protect\citeauthoryear{Moser, Pedrycz, and Succi}{Moser
  et~al\mbox{.}}{2008}]%
        {Moser2008}
\bibfield{author}{\bibinfo{person}{Raimund Moser}, \bibinfo{person}{Witold
  Pedrycz}, {and} \bibinfo{person}{Giancarlo Succi}.}
  \bibinfo{year}{2008}\natexlab{}.
\newblock \showarticletitle{{A Comparative analysis of the efficiency of change
  metrics and static code attributes for defect prediction}}.
\newblock \bibinfo{journal}{\emph{Proceedings - International Conference on
  Software Engineering}} \bibinfo{number}{January 2008} (\bibinfo{year}{2008}),
  \bibinfo{pages}{181--190}.
\newblock
\showISBNx{9781605580791}
\showISSN{02705257}
\urldef\tempurl%
\url{https://doi.org/10.1145/1368088.1368114}
\showDOI{\tempurl}


\bibitem[\protect\citeauthoryear{Nagappan and Ball}{Nagappan and Ball}{2005}]%
        {Nagappan2005}
\bibfield{author}{\bibinfo{person}{Nachiappan Nagappan} {and}
  \bibinfo{person}{Thomas Ball}.} \bibinfo{year}{2005}\natexlab{}.
\newblock \showarticletitle{{Use of relative code churn measures to predict
  system defect density}}.
\newblock \bibinfo{journal}{\emph{Proceedings - 27th International Conference
  on Software Engineering, ICSE05}} (\bibinfo{year}{2005}),
  \bibinfo{pages}{284--292}.
\newblock
\showISBNx{1581139632}
\urldef\tempurl%
\url{https://doi.org/10.1145/1062455.1062514}
\showDOI{\tempurl}


\bibitem[\protect\citeauthoryear{Nam, Fu, Kim, Menzies, and Tan}{Nam
  et~al\mbox{.}}{2018}]%
        {Nam2018}
\bibfield{author}{\bibinfo{person}{Jaechang Nam}, \bibinfo{person}{Wei Fu},
  \bibinfo{person}{Sunghun Kim}, \bibinfo{person}{Tim Menzies}, {and}
  \bibinfo{person}{Lin Tan}.} \bibinfo{year}{2018}\natexlab{}.
\newblock \showarticletitle{{Heterogeneous Defect Prediction}}.
\newblock \bibinfo{journal}{\emph{IEEE Transactions on Software Engineering}}
  \bibinfo{volume}{44}, \bibinfo{number}{9} (\bibinfo{year}{2018}),
  \bibinfo{pages}{874--896}.
\newblock
\showISSN{19393520}
\urldef\tempurl%
\url{https://doi.org/10.1109/TSE.2017.2720603}
\showDOI{\tempurl}


\bibitem[\protect\citeauthoryear{Pascarella, Palomba, and Bacchelli}{Pascarella
  et~al\mbox{.}}{2019}]%
        {Pascarella2019}
\bibfield{author}{\bibinfo{person}{Luca Pascarella}, \bibinfo{person}{Fabio
  Palomba}, {and} \bibinfo{person}{Alberto Bacchelli}.}
  \bibinfo{year}{2019}\natexlab{}.
\newblock \showarticletitle{{Fine-grained just-in-time defect prediction}}.
\newblock \bibinfo{journal}{\emph{Journal of Systems and Software}}
  \bibinfo{volume}{150} (\bibinfo{year}{2019}), \bibinfo{pages}{22--36}.
\newblock
\showISSN{01641212}
\urldef\tempurl%
\url{https://doi.org/10.1016/j.jss.2018.12.001}
\showDOI{\tempurl}


\bibitem[\protect\citeauthoryear{Qiao and Wang}{Qiao and Wang}{2019}]%
        {Qiao2019}
\bibfield{author}{\bibinfo{person}{Lei Qiao} {and} \bibinfo{person}{Yan Wang}.}
  \bibinfo{year}{2019}\natexlab{}.
\newblock \showarticletitle{{Effort-aware and just-in-time defect prediction
  with neural network}}.
\newblock \bibinfo{journal}{\emph{PLoS ONE}} \bibinfo{volume}{14},
  \bibinfo{number}{2} (\bibinfo{year}{2019}), \bibinfo{pages}{1--19}.
\newblock
\showISSN{19326203}
\urldef\tempurl%
\url{https://doi.org/10.1371/journal.pone.0211359}
\showDOI{\tempurl}


\bibitem[\protect\citeauthoryear{{\'{S}}liwerski, Zimmermann, and
  Zeller}{{\'{S}}liwerski et~al\mbox{.}}{2005}]%
        {Sliwerski2005}
\bibfield{author}{\bibinfo{person}{Jacek {\'{S}}liwerski},
  \bibinfo{person}{Thomas Zimmermann}, {and} \bibinfo{person}{Andreas Zeller}.}
  \bibinfo{year}{2005}\natexlab{}.
\newblock \showarticletitle{{When do changes induce fixes?}}
\newblock \bibinfo{journal}{\emph{Proceedings of the 2005 International
  Workshop on Mining Software Repositories, MSR 2005}} (\bibinfo{year}{2005}).
\newblock
\showISBNx{1595931236}
\showISSN{01635948}
\urldef\tempurl%
\url{https://doi.org/10.1145/1082983.1083147}
\showDOI{\tempurl}


\bibitem[\protect\citeauthoryear{Son, Pritam, Khari, Kumar, Phuong, and
  Thong}{Son et~al\mbox{.}}{2019}]%
        {Son2019}
\bibfield{author}{\bibinfo{person}{Le~Hoang Son}, \bibinfo{person}{Nakul
  Pritam}, \bibinfo{person}{Manju Khari}, \bibinfo{person}{Raghvendra Kumar},
  \bibinfo{person}{Pham Thi~Minh Phuong}, {and} \bibinfo{person}{Pham~Huy
  Thong}.} \bibinfo{year}{2019}\natexlab{}.
\newblock \showarticletitle{{Empirical study of software defect prediction: A
  systematic mapping}}.
\newblock \bibinfo{journal}{\emph{Symmetry}} \bibinfo{volume}{11},
  \bibinfo{number}{2} (\bibinfo{year}{2019}).
\newblock
\showISSN{20738994}
\urldef\tempurl%
\url{https://doi.org/10.3390/sym11020212}
\showDOI{\tempurl}


\bibitem[\protect\citeauthoryear{Wang and Li}{Wang and Li}{2010}]%
        {Wang2010}
\bibfield{author}{\bibinfo{person}{Tao Wang} {and} \bibinfo{person}{Wei~Hua
  Li}.} \bibinfo{year}{2010}\natexlab{}.
\newblock \showarticletitle{{Na{\"{i}}ve bayes software defect prediction
  model}}.
\newblock \bibinfo{journal}{\emph{2010 International Conference on
  Computational Intelligence and Software Engineering, CiSE 2010}}
  \bibinfo{number}{2006} (\bibinfo{year}{2010}), \bibinfo{pages}{1--4}.
\newblock
\showISBNx{9781424453924}
\urldef\tempurl%
\url{https://doi.org/10.1109/CISE.2010.5677057}
\showDOI{\tempurl}


\bibitem[\protect\citeauthoryear{{Xu-Ying Liu}, {Jianxin Wu}, and {Zhi-Hua
  Zhou}}{{Xu-Ying Liu} et~al\mbox{.}}{2009}]%
        {Xu-YingLiu2009}
\bibfield{author}{\bibinfo{person}{{Xu-Ying Liu}}, \bibinfo{person}{{Jianxin
  Wu}}, {and} \bibinfo{person}{{Zhi-Hua Zhou}}.}
  \bibinfo{year}{2009}\natexlab{}.
\newblock \showarticletitle{{Exploratory Undersampling for Class-Imbalance
  Learning}}.
\newblock \bibinfo{journal}{\emph{IEEE Transactions on Systems, Man, and
  Cybernetics, Part B (Cybernetics)}} \bibinfo{volume}{39}, \bibinfo{number}{2}
  (\bibinfo{date}{apr} \bibinfo{year}{2009}), \bibinfo{pages}{539--550}.
\newblock
\showISBNx{1083-4419}
\showISSN{1083-4419}
\urldef\tempurl%
\url{https://doi.org/10.1109/TSMCB.2008.2007853}
\showDOI{\tempurl}
\showeprint[arxiv]{arXiv:1011.1669v3}


\bibitem[\protect\citeauthoryear{Yan, Fang, Lo, Xia, and Zhang}{Yan
  et~al\mbox{.}}{2017}]%
        {Yan2017}
\bibfield{author}{\bibinfo{person}{Meng Yan}, \bibinfo{person}{Yicheng Fang},
  \bibinfo{person}{David Lo}, \bibinfo{person}{Xin Xia}, {and}
  \bibinfo{person}{Xiaohong Zhang}.} \bibinfo{year}{2017}\natexlab{}.
\newblock \showarticletitle{{File-Level Defect Prediction: Unsupervised vs.
  Supervised Models}}.
\newblock \bibinfo{journal}{\emph{International Symposium on Empirical Software
  Engineering and Measurement}}  \bibinfo{volume}{2017-Novem}
  (\bibinfo{year}{2017}), \bibinfo{pages}{344--353}.
\newblock
\showISBNx{9781509040391}
\showISSN{19493789}
\urldef\tempurl%
\url{https://doi.org/10.1109/ESEM.2017.48}
\showDOI{\tempurl}


\bibitem[\protect\citeauthoryear{Yang, Zhou, Liu, Zhao, Lu, Xu, Xu, and
  Leung}{Yang et~al\mbox{.}}{2016}]%
        {Yang2016}
\bibfield{author}{\bibinfo{person}{Yibiao Yang}, \bibinfo{person}{Yuming Zhou},
  \bibinfo{person}{Jinping Liu}, \bibinfo{person}{Yangyang Zhao},
  \bibinfo{person}{Hongmin Lu}, \bibinfo{person}{Lei Xu},
  \bibinfo{person}{Baowen Xu}, {and} \bibinfo{person}{Hareton Leung}.}
  \bibinfo{year}{2016}\natexlab{}.
\newblock \showarticletitle{{Effort-Aware just-in-Time defect prediction:
  Simple unsupervised models could be better than supervised models}}.
\newblock \bibinfo{journal}{\emph{Proceedings of the ACM SIGSOFT Symposium on
  the Foundations of Software Engineering}}  \bibinfo{volume}{13-18-Nove}
  (\bibinfo{year}{2016}), \bibinfo{pages}{157--168}.
\newblock
\showISBNx{9781450342186}
\urldef\tempurl%
\url{https://doi.org/10.1145/2950290.295035}
\showDOI{\tempurl}


\end{thebibliography}

\end{document}